\documentclass[article,showpacs,floats,floatfix,nofootinbib, usenatbib]{mn2e}

\usepackage{amsbsy}
\usepackage{amssymb}
\usepackage{amsmath}
\input{epsf}
\usepackage{graphicx}
\usepackage{color}

\usepackage{amssymb}
\usepackage{bm}
\def\spose#1{\hbox to 0pt{#1\hss}}

\def\lta{\mathrel{\spose{\lower 3pt\hbox{$\mathchar"218$}}
     \raise 2.0pt\hbox{$\mathchar"13C$}}}
\def\gta{\mathrel{\spose{\lower 3pt\hbox{$\mathchar"218$}}
     \raise 2.0pt\hbox{$\mathchar"13E$}}}

\def\beq{\begin{equation}}
\def\eeq{\end{equation}}
\def\bea{\begin{eqnarray}}
\def\eea{\end{eqnarray}}

\begin{document}
\title{Exploring universality in neutron star mergers}

\author[K. Chakravarti \& N. Andersson]{Kabir Chakravarti$^1$ and Nils Andersson$^2$\\
$^1$ IUCAA, Ganeshkhind, Post Bag 4, Pune University Campus, Pune 411007, India; email: kabir@iucaa.in\\
$^2$ Mathematical Sciences and STAG Research Centre, University of Southampton, Southampton 
SO17 1BJ, UK}

\date{\today}

\maketitle

\begin{abstract}
We explore the correlation between the pre-merger tidal deformability and the post-merger remnant oscillations seen in numerical simulation of neutron star binaries, with the aim of understanding to what extent the physics support the existence of such a relation. We consider the impact of thermal and rotational effects and argue that the proposed relation does, indeed, make sense and provide simple arguments that help explain the result.   
\end{abstract}

\section{Introduction}

Gravitational waves from the inspiral and merger of binary neutron stars encode the extreme physics of these objects. From the inspiral phase one may extract the faint imprint of the tidal interaction while the violent merger involves the oscillations of the hot  remnant that may eventually collapse to form a black hole. The first aspect has already been demonstrated in the case of the spectacular GW170817 signal \citep{TheLIGOScientific:2017qsa}. The contribution from the subsequent merger, which leaves a  higher frequency signature, was however buried in the instrument noise in this instance \citep{Abbott:2017dke,Abbott:2018wiz}.  Intuitively, there is no reason why these distinct phases of the gravitational-wave signal should be correlated. Yet, numerical simulations suggest we should expect the unexpected.

A typical neutron star merger leads to the formation of a massive, $3-4 M_{\odot}$,  remnant  (often referred to as a Hyper Massive Neutron Star, HMNS), which collapses to a black hole (on a timescale of 100~ms) after shedding the angular momentum that (temporarily) counteracts gravity \citep{Takami:2014tva}. Numerical simulations  demonstrate that the dynamics of the HMNS have robust high-frequency features \citep{Bauswein:2015vxa,Rezzolla:2016nxn}, thought to be associated with the modes of oscillation of the remnant. This would be natural, although the problem is less straightforward than the corresponding one for a cold neutron star  (see for example \citet{Cunha:2007wx} or \citet{Chirenti:2012wn}). In that case one would be considering perturbations with respect to a long-term stable background. Meanwhile, in the case of a HMNS one would have  to explore perturbations relative to a background that evolves (and eventually collapses) on a relatively short timescale.  The intuitive picture makes sense, but we do not (at least not yet)  have precise mode-calculations to test simulations (and eventually observations) against.

Numerical simulations have demonstrated the existence of useful phenomenological relations linking the post-merger oscillation frequencies to the matter equation of state \citep{,Bauswein:2012ya,Rezzolla:2016nxn}. This is important as it means that observations could eventually help us get a handle on problematic physics associated with hot high-density matter. This information would complement information gleaned from the inspiral phase (e.g. in terms of the tidal deformability, see \citet{Flanagan:2007ix} and \citet{Hinderer:2007mb}), which relates to cold supranuclear matter. However, it is not clear to what extent the information from inspiral and merger is  (at the end of the day)  independent. Formally, the underlying equation of state should (obviously) be the same (involving identical many-body interactions etcetera) but one might expect thermal and rotational effects to have decisive impact on the HMNS. Given this, it is interesting to note that
the tidal deformability (usually encoded in a mass-weighted combination of the so-called Love numbers of the individual stars, $\kappa_2^t$ in the following \citep{Flanagan:2007ix}) appears to be linked to the dominant oscillation frequency ($f_2$) of the post-merger remnant, see in particular \citet{Bernuzzi:2015rla}. The relation appears to be robust, perhaps hinting at some underlying universality, and may provide a useful constraint on the inferred physics.  Of course, before we make use of this information in either a data analysis algorithm or a parameter extraction effort we need to understand why the relation should hold. At first sight it seems peculiar. Why should the properties of the (cold, slowly spinning) inspiralling neutron stars be related to the oscillations of the (hot, differentially rotating) remnant? This is the question we (try to) address in the following.


\section{The implied universality}

In the last few years it has become clear that many neutron star properties are related through universal relations. In fact, since the late 1990s, this observation has provided a foundation for discussions of neutron star asteroseismology \citep{Andersson:1997rn}, which aims to use observed oscillation frequencies to infer mass and radius (and hence constrain the equation of state) for individual stars. More recently, the so-called I-Love-Q relations \citep{Yagi:2013bca} demonstrate a  link between the moment of inertia, the Love number and the quadrupole moment  that helps break degeneracies in gravitational-waveform modelling. Finally, one may link the f-mode frequency of a given star to the tidal deformability \citep{Chan:2014kua}. Since these relations have been demonstrated to be accurate to within a few percent (at least as long as the equation of state does not involve sharp phase transitions, see \citet{han}) it makes sense to take them as our starting point. 

Labelling the binary partners $a$ and $b$, with the mass ratio $q = {M_b}/{M_a} \leq 1$, the effective tidal parameter used by, for example,  \citet{Bernuzzi:2015rla} is given by
\begin{equation}
\kappa_2^t = 2\left[ q \left(\frac{X_a}{C_a}\right)^5 k_2^a+
\frac{1}{q} \left(\frac{X_b}{C_b}\right)^5k_2^b \right] 
\end{equation}
where $C_a=M_a/R_a$ is the compactness of each star,  $X_a= {M_a}/{(M_a+M_b)}$ while $k_2^a$ is the quadrupole Love number (and similarly for star $b$).
For simplicity, let us consider a binary system with two non-spinning equal-mass partners. Then we have
\begin{equation}
\kappa_2^t = \frac{1}{8} \frac{k_2}{C^5} = {3\over 16} \Lambda
\end{equation}
where $\Lambda$ is commonly used to quantify the tidal deformability.
If, in addition, we note that $k_2$ is only weakly dependent on the compactness, we expect the scaling
\begin{equation}
\kappa_2^t \sim C^{-5}
\end{equation}

Meanwhile, we know that (for non-spinning stars) the fundamental mode frequency scales (roughly) as the average density (see for example \citet{Andersson:1997rn}). That is, we have 
\begin{equation}
Mf_2 \sim M\bar{\rho}^{1/2} = C^{3/2} 
\end{equation}
In essence, for a cold neutron star we expect to have
\begin{equation}  \label{scaling}
Mf_2 \sim C^{3/2} \sim \left(\kappa_2^t\right)^{-3/10}
\end{equation}

\begin{figure}
\centerline{\includegraphics[width=0.4\textwidth]{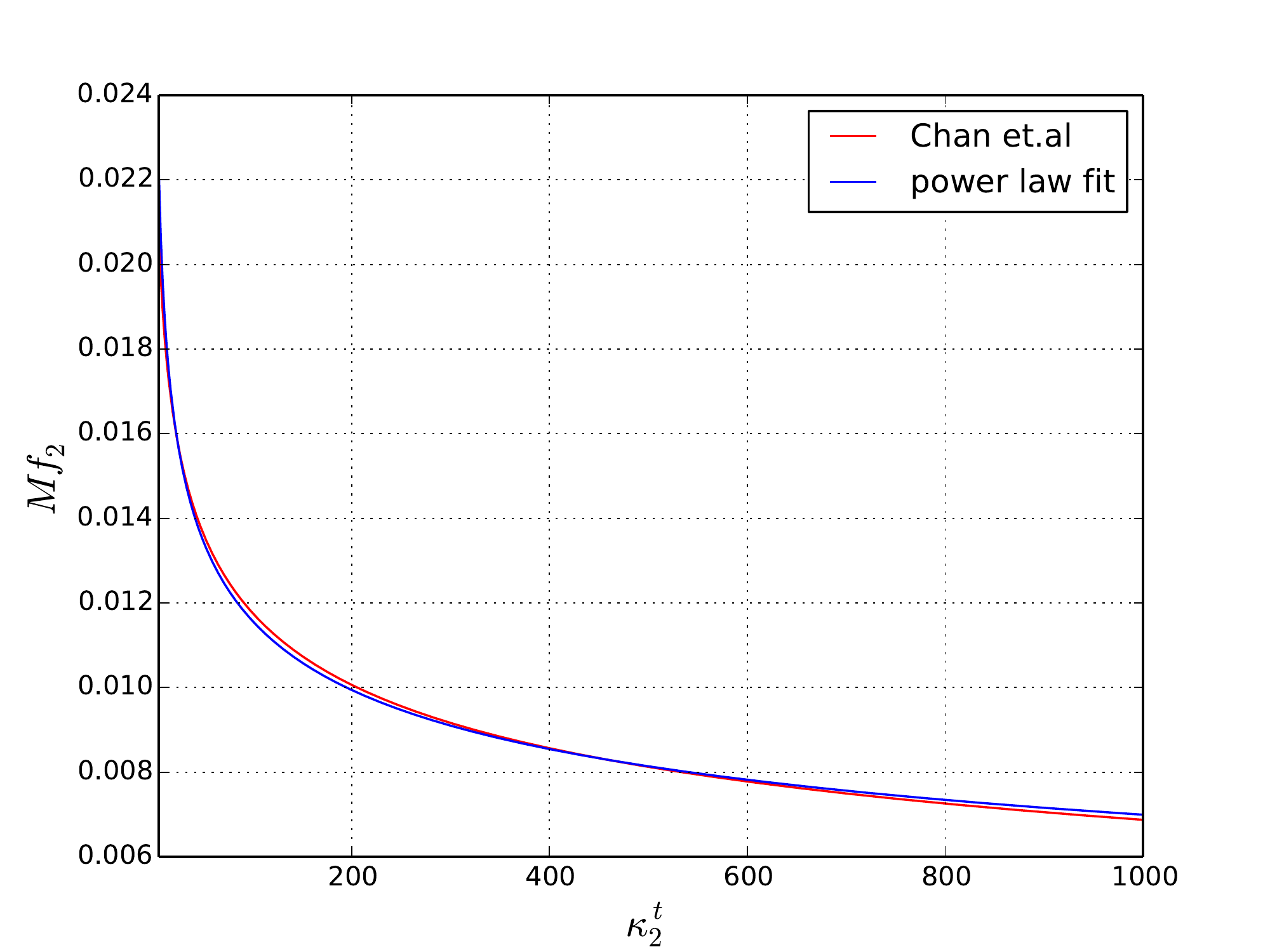}}
\caption{Power-law fit to the $Mf_2-\kappa_2^t$ relation for cold neutron stars from \citet{Chan:2014kua}, demonstrating that we can reliably base the discussion on the simpler scaling from \eqref{f-mode_premerg}.}
\label{gchan}
\end{figure}

This is, of course, only a rough indication.  A more precise relation between the f-mode frequency and the tidal deformability has been obtained by \citet{Chan:2014kua}, linking the dimensionless frequency (in geometric units) $\omega M$ to an expansion in $\ln \Lambda$. This relation is a little bit too complicated for our purposes, but it is easy to demonstrate that it can be replaced by the  power-law
\begin{equation}\label{f-mode_premerg}
Mf_2 \approx 0.031  \left(\kappa_2^t\right)^{-0.218} \ .
\end{equation}
Moreover, as is evident from figure~\ref{gchan}, this is  is close to the $-3/10$ power law suggested by our simple argument.  Basically, the origin of the scaling for cold neutron stars is well understood.

\begin{figure}
\centerline{\includegraphics[width=0.4\textwidth]{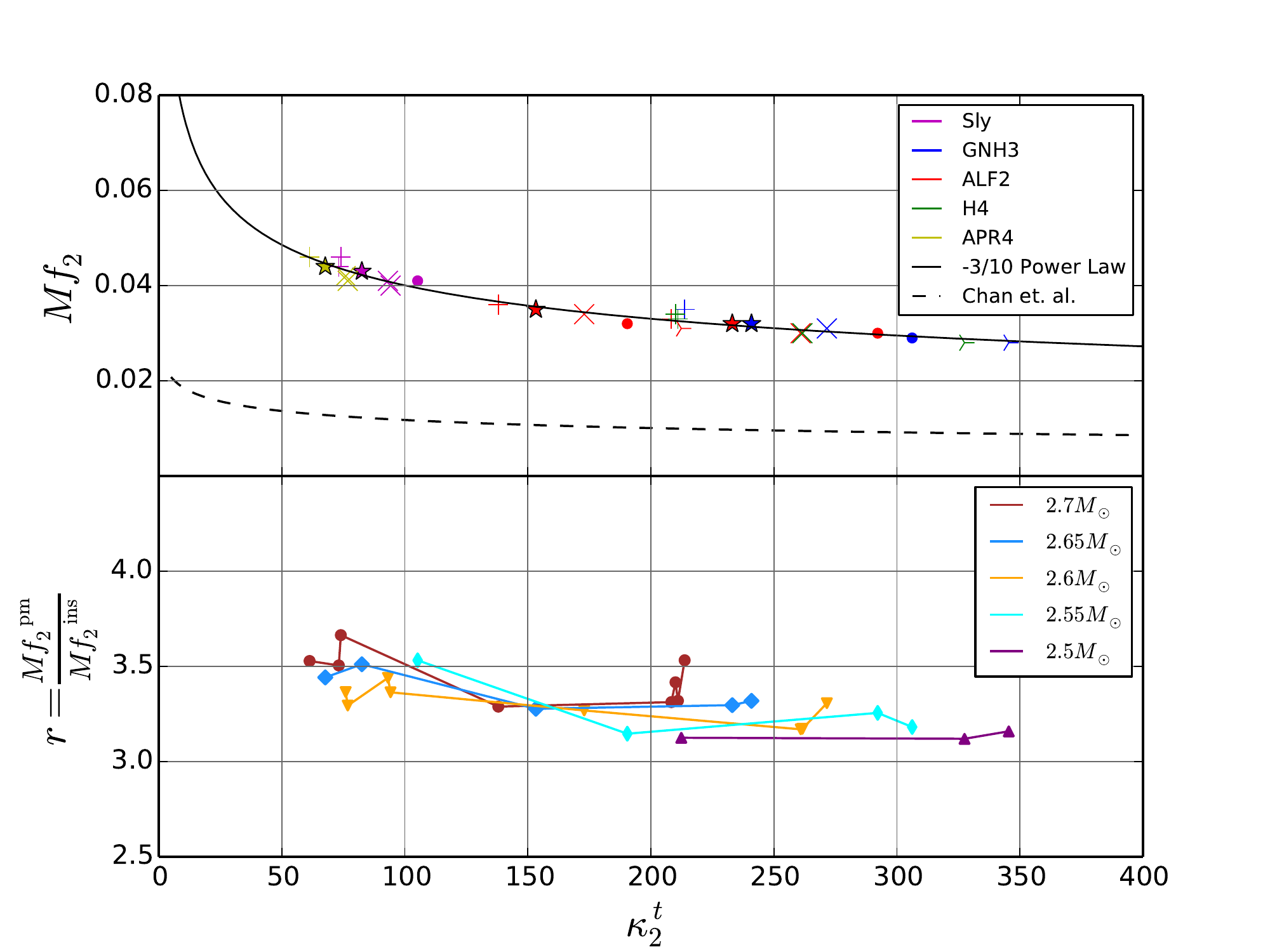}}
\caption{The top panel shows the suggested $-3/10$ power-law fit to the post-merger f-modes inferred from simulations for a set of equations of state (solid line, data provided by S. Bernuzzi). Also indicated (as a dashed line)  is the fit to the f-mode of the individual (cold) pre-merger neutron stars  from \citet{Chan:2014kua}. The bottom panel illustrates the fractional increase in the f-mode frequency required to explain the scaling of the post-merger remnant oscillations (assuming that no mass is lost during the merger). This factor is seen lie in the range $3-4$.  }
\label{f2vk}
\end{figure}

This is not the scaling relation we are interested in, but it is easy to show that the oscillations of the remnant are well represented by a power law, as well (see \citet{Takami:2014tva} and \citet{Rezzolla:2016nxn}). Using data from a range of simulations (by different groups) we infer the scaling (see figure~\ref{f2vk})
\begin{equation}\label{f-mode_postmerg}
 M_tf_2^h \approx 0.144  (\kappa_2^t)^{-0.278},
\end{equation}
where $M_t$ is the total mass of the system (assuming that we can neglect mass shedded during the merger) and $f_2^h$ represents the hot post-merger f-mode. Notably, the post-merger f-modes follow almost exactly the same scaling law as the cold f-mode of the individual pre-merger stars. This is the behaviour we are trying to explain. 

The problem breaks down to two questions. First of all, we need to understand why ``the same'' scaling with the tidal parameter should apply for the oscillations of cold neutron stars and hot, more massive and differentially rotating, merger remnants. Secondly, we note that the scalings \eqref{f-mode_premerg} and \eqref{f-mode_postmerg} imply that 
\begin{equation}\label{beta}
\left(M_t f_2^h\right) \approx \beta \left(Mf_2\right) 
\end{equation}
with $\beta \approx 3-4$ (see figure~\ref{f2vk}). We  need (at least at the qualitative level) to understand how the merger physics impacts on this numerical factor. 

\section{Thermal Effects}

Just after the merger, the HMNS can reach temperatures as high as 85~MeV \citep{, Hanauske:2016gia,Hanauske:2019qgs}, making the remnant hotter than the collapsing core of a supernova furnace. Given this, one would expect thermal effect to come into play and we need to explore the implicates for the f-mode oscillations. However, the problem is complicated by the fact that the system is evolving -- we are not dealing with a ``thermalized'' equlibrium background with respect to which we can define a mode perturbation. Still, we can make progress with a simple argument.  As we have already pointed out, the f-modes  are known to be determined by the average density of the body in question. For the HMNS, this is tricky, as $\bar{\rho}$ (or equivalently, as the mass is known, the compactness $C$) of the collapsing matter is explicitly dependent on time. However, as the suggested collapse timescale ($\thicksim 0.1$ s) is about two orders of magnitude larger than the f-mode oscillation timescale (typically $\thicksim 1$ ms) we can expect the HMNS density to be roughly constant on the timescale of a few oscillations. This is all we need to progress, as it allows us to simplify the discussion by considering a single ``neutron star''. However, it is not quite enough to make the argument quantitative. After all,  stable solutions can never reach masses of $3-4M_{\odot}$. This likely requires both support of thermal pressure and differential rotation (see later). Nevertheless, we may be able to gain some insight into the thermal effects from a simple ``surrogate'' model. Basically, we should be able to estimate the effect on the f-mode as the star becomes bloated due to thermal pressure, causing a relative change in the average density and the compactness. This effect has, in fact, already been studied for proto-neutron stars \citep{Ferrari:2002ut}. The main difference here is that we consider a model inspired by the actual temperature distribution inside a HMNS.

The thermal profile of a HMNS is not uniform. In essence, the hottest regions arise from heating due to shocks as the stars come into contact. The remnant is not able to to reach thermal equilibrium on the timescale we are interested in. Instead, the heat is advected along with the matter, leading to a nonuniform temperature distribution, see for example the results of \citet{Hanauske:2016gia}. In general, the thermal distribution has no obvious symmetry. However, for simplicity we will assume that it has a simple radial profile. Motivated by Figure 6 of \citet{Hanauske:2016gia} we assume the temperature profile to be such that:

\begin{itemize}
\item[-] thermal effects can be ignored up to about a 5~km distance from centre, simply because the pressure of the cold high-density matter dominates in this region,

\item[-] the temperature reaches 50~MeV temperature in the region between 5-10~km,

\item[-] the temperature drops to about 10~MeV temperature in the region between 10-15~km,

\item[-] we (again) ignore thermal effects beyond a radius of 15~km as the temperature (indicated by simulations) is close to zero (and this matter is likely to be gravitationally unbound, anyway). 
\end{itemize} 

With this model as our starting point, we can easily calculate the relative compactness of hot and cold models. This, in turn, provides some insight into the thermal impact on the post-merger oscillations. In effect, the  post-merger (hot) compactness can be related to the pre-merger (cold) compactness which then relates to the tidal deformability in terms of  $\kappa_2^t$. 

In order to estimate the magnitude of this effect, which obviously leads to different fluid configurations, we need some idea of how the compactness of the merger remnant differs from that of the original cold neutron star. This is not at all trivial, but   we may try to encode the change  in compactness in terms of a nearly constant factor, $F(\theta,\phi)$. Then we argue that we ought to have $F\leq 2$, following from taking the (equatorial) radius at the point when the stars first touch. Meanwhile, numerical simulations  show that we should have $F \geq 1$. For simplicity, we ignore the $\theta,\phi$ dependence which relates to the deviation from spherical symmetry. If we now imagine a hypothetical postmerger compactness,  $C_{hyp}$,  we have
\begin{equation}\label{chyp1}
C_{hyp} = F(\theta,\phi) C
\end{equation} 
However, we still need to quantify the  thermal effects.

\begin{figure}
\centerline{\includegraphics[width=0.4\textwidth]{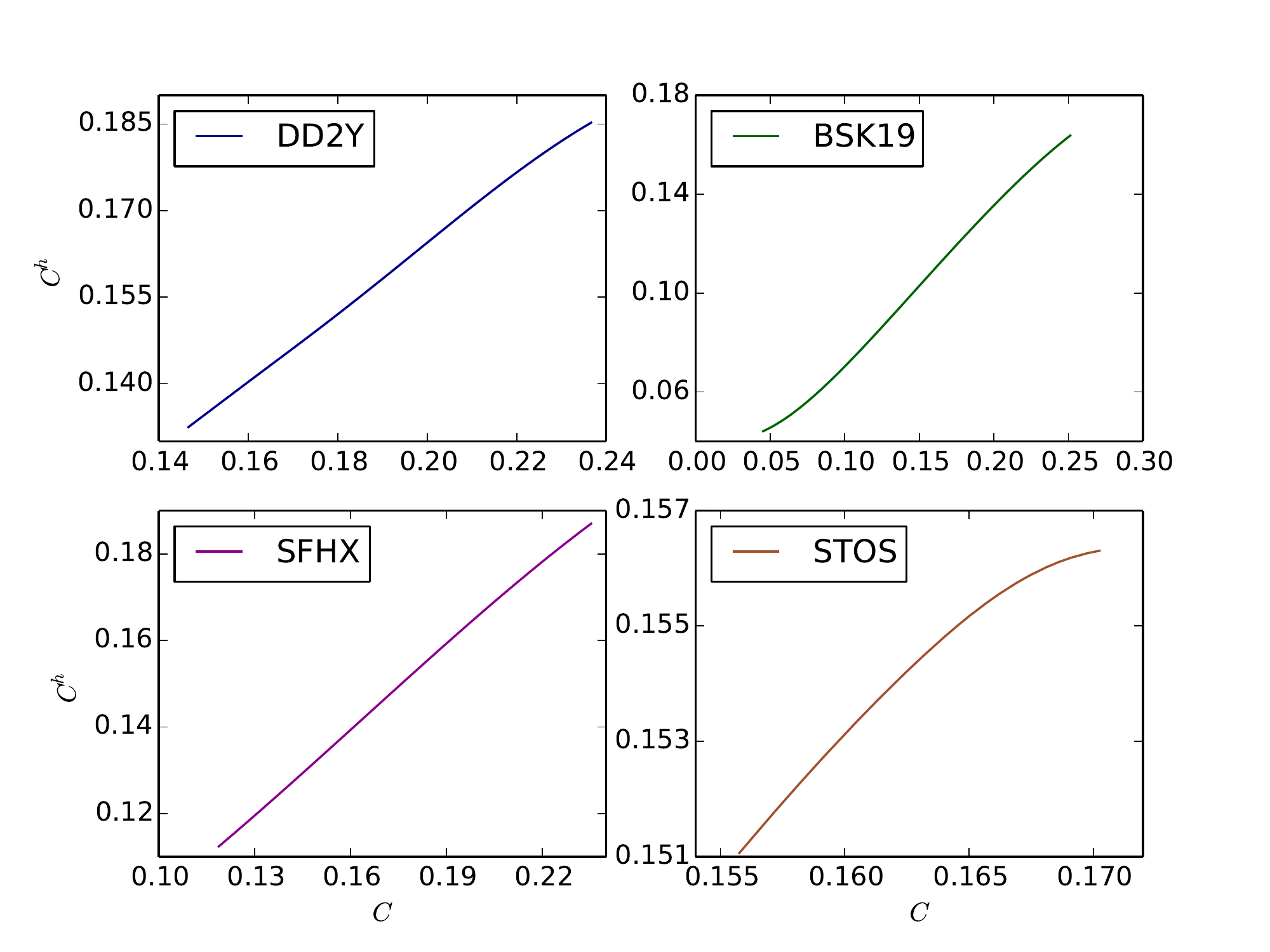}}
\caption{Illustrating the relation between  the ``hot'' compactness $C^h$  (for our simple thermal surrogate model) and the cold compactness $C$ for for four different equations of state (EoS).}
\label{hotvcold}
\end{figure}

Results from such an  exercise are shown in figure~\ref{hotvcold}, which shows the relation between the ``hot'' compactness  and the corresponding cold model, based on the thermal surrogate and four different EoSs. It is notable that the relationship is close to linear for a range of parameter values. This is true for all models we have considered. Since the thermal surrogate ought to capture the relative behaviour, we therefore expect the postmerger compactness $C^h$ to scale linearly with $C_{hyp}$. That is, we have
\begin{equation}\label{chyp2}
C^h = \alpha C_{hyp} \quad \alpha <1
\end{equation}
Combining \eqref{chyp1} and \eqref{chyp2}, we finally obtain
\begin{equation}\label{comp_eq}
C^h = (\alpha F) C ,
\end{equation}
where the pre-factors are expected to be roughly constant. This is an important conclusion. It is now apparent why the thermal pressure would not affect the power-law from \eqref{scaling}. 

Moreover, we can estimate the quantitative effect on the f-mode frequency. The results in figure~\ref{hotvcold} suggest that  $\alpha\approx 0.6-0.9$, leading to the estimated range $F\approx 1.5-1.7$. In essence, we expect the post-merger compactness to be  $1 -1.5$ times the pre-merger value. The post-merger frequencies would then increase by factor of $(1-1.5)^{3/2}$ or $\approx(1 - 2)$ because of the  thermal effects, which takes us some way towards explaining the missing factor indicated by the results in figure~\ref{f2vk}. However, we still seem to be short a factor of 2 or so. 

Before we move on to consider whether rotation may provide the missing factor, let us consider a related question.
In a series of papers,  see for example \citet{Bauswein:2011tp} and \citet{Bauswein:2015vxa}, it has been argued that the f-mode frequency of the HMNS scales with the radius of an isolated neutron star with mass $1.6M_{\odot}$. This scaling is interesting because it is not at all obvious why the f-mode of the hot, differentially rotating, star should depend on the radius of an isolated neutron star with a particular mass. Further, the scaling seems to be insensitive to the mass of the HMNS.

Motivated by this, we break down the argument for the  observed scaling of $M_tf_2^h$ with the isolated neutron star mass into three steps.  First of all, given a particular remnant there must exist a variation of the radius of the remnant across different EoS which is unaffected by the mass of the remnant. If this is the case, we  can calculate the mass of an isolated NS that reproduces this variation in radius.
Finally, since $M_tf_2^h \thicksim \left(C^h\right)^{-3/2} \thicksim \left(R^h\right)^{-3/2}$ the f-mode should scale with the radius  of the isolated neutron star for which the radius varies with EoS in the same way as $R^h$.

\begin{figure}
\centerline{\includegraphics[width=0.4\textwidth]{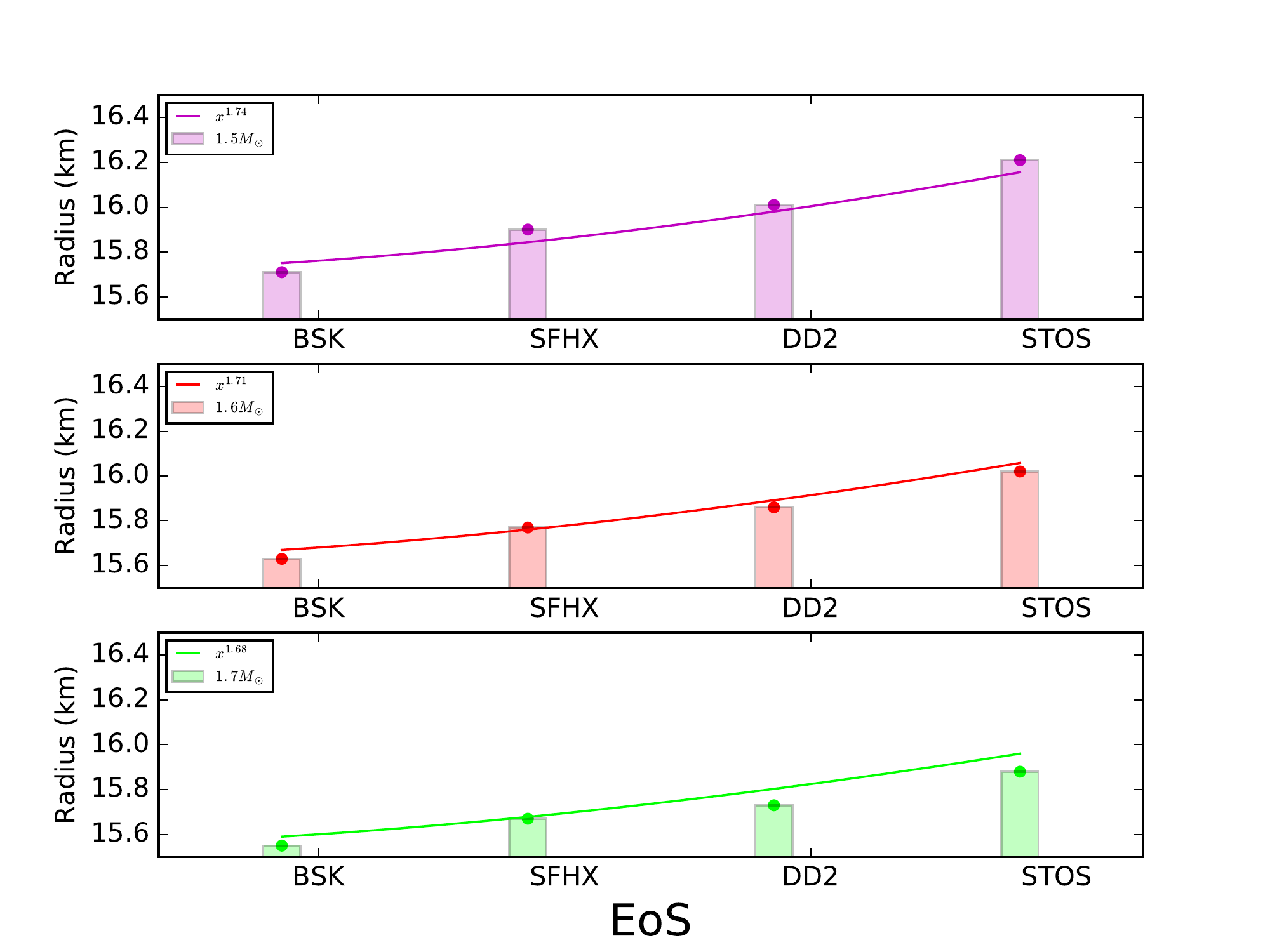}}
\caption{Variation of  the ``hot'' neutron star radius for different EoS (for the thermal surrogate model) and three different masses. The fits show  power laws that best capture the variation, having first of all ordered the equations of state in order of increasing radius (for stars of mass $1.5M_\odot$) and then compared the scaling (in terms of a fiducial equation of state parameter $x$) for different masses. It is notable these power laws are similar ($\sim 1.7$) for masses in the range $1.5-1.7M_\odot$. }
\label{hotradius}
\end{figure}

\begin{figure}
\centerline{\includegraphics[width=0.4\textwidth]{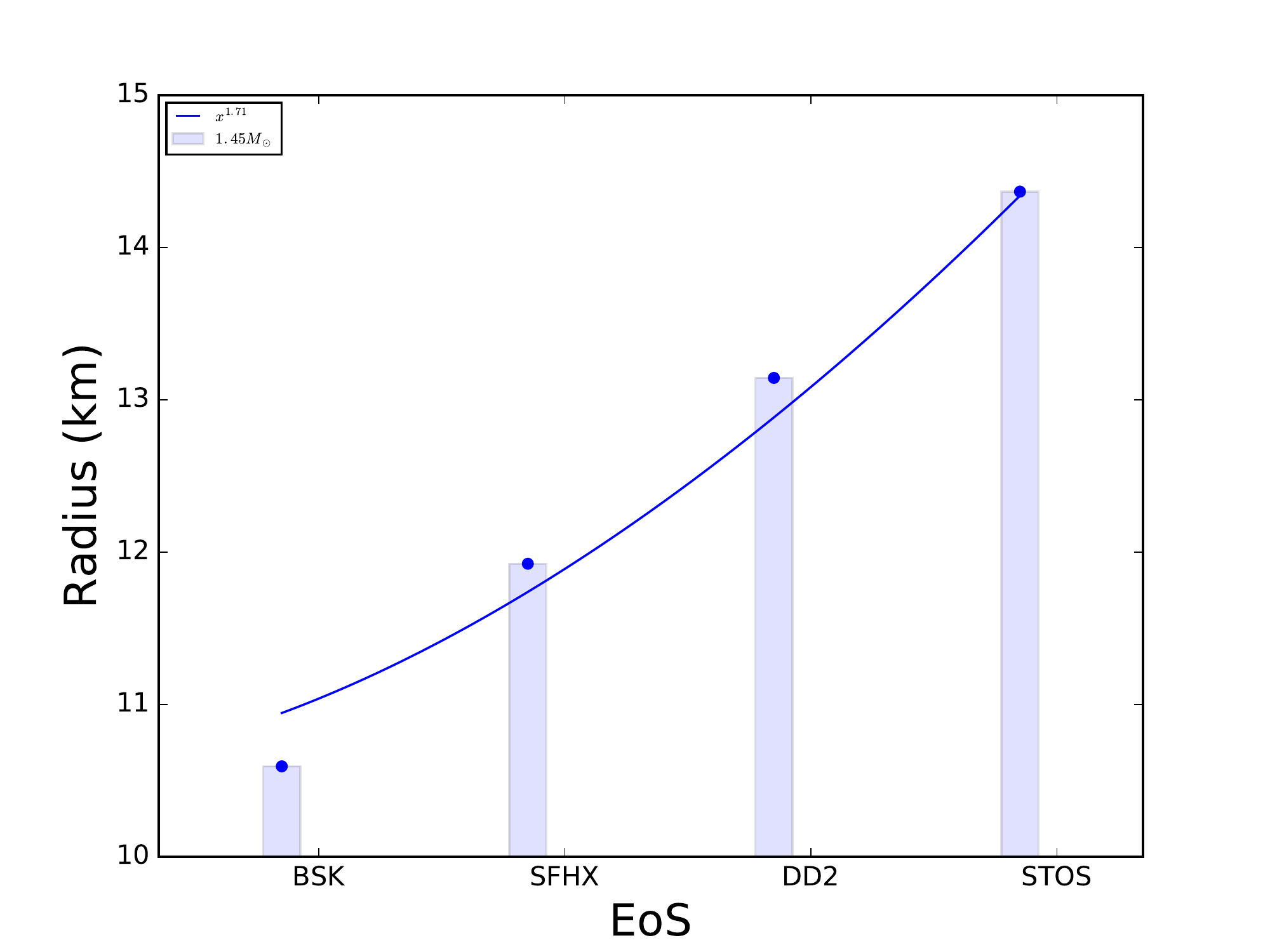}}
\caption{Variation of radius with EoS for $1.45 M_{\odot}$ isolated neutron star models. The indicated power-law of $\approx 1.7$ is that required by the results in figure~\ref{hotradius}.}
\label{coldradius}
\end{figure}

Once again we turn to the thermal surrogate model for an answer. Figure~\ref{hotradius} shows the variation of the radius of the surrogate models  for three different masses and different EoS. As before, we emphasize that the surrogate model will not return to ``actual'' value of the radius, but we expect that with the EoS will be meaningfully captured.  It is then clear from figure~\ref{hotradius} that the variation of radii across EoS closely follows a 1.7 power law. This completes the first item on our list. This moves us on to the issue of determining a mass value for an isolated neutron star showing the same variation. This  solution to this problem is  provided in figure~\ref{coldradius}. The results suggest that the f-modes should scale with the radius of a $1.45 M_\odot$ neutron star, not too different from the  $1.6 M_\odot$ scaling of  \citet{Bauswein:2015vxa} (especially if we consider that we are using a fairly crude argument).

\section{Rotational effects} 

Let us turn to the role of rotation. As in the thermal case, it is easy to argue that the rapid rotation of the HMNS will have decisive impact on the f-mode oscillation frequencies. In fact, in this case we have better quantitative evidence. We may, for example, draw on the results from \citet{Doneva:2013zqa}, which show how the f-mode of an isolated NS changes with the angular frequency $\Omega$. The key point is that there exists robust phenomenological relations drawn from a collection of EoS. These support the notion that the overall scaling with rotation is likely to be insensitive to the EoS, as required to explain the results in figure~\ref{f2vk}. However, we have to be a bit careful as the results assume uniform rotation, while we know that the HMNS will rotate differentially. At the same time, the typical rotation profile inferred from simulations \citep{uryu} has the high-density core rotating close to uniformly. The argument may be strengthened by mode calculations for the appropriate HMNS differential rotation law, but such work has not yet been carried out. 

With this caveat in mind, let us piece together the argument from the results from  \citet{Doneva:2013zqa}. In order to do this, we need some rough idea of the spin of the HMNS. This is relatively straightforward. Assuming that the individual stars in the binary are slowly rotating (which makes sense if the system is old enough that the stars have had time to spin down due to dipole emission, which is likely), the angular velocity of the HMNS should arise from the total angular momentum of the system at the innermost stable circular orbit.  Conservation of angular momentum then allows us to estimate the HMNS rotation rate. This rough estimate agrees fairly well with the results inferred from simulations, which lead to a dimensionless rotation parameter
\begin{equation}
\frac{a}{M} = \frac{J}{M^2} = \tilde{I} \left(\frac{R}{M}\right)^2 (M\Omega) \thicksim (0.75-0.8)
\end{equation}
We can turn this into an estimate for the (here assumed uniform) rotation via the scaling for the dimensionless moment of inertia, $\bar{I}$, from \citet{Lattimer:2004nj}
\begin{equation}
\tilde{I} = 0.237\left[ 1 + 4.2\frac{M/M_{\odot}}{R/\mathrm{km}} + 90 \left(\frac{M/M_{\odot}}{R/\mathrm{km}}\right)^4  \right]
\end{equation}
Working out the rotation rate from these relations  for (say) a  remnant mass, $2.6 M_{\odot}$ with 15~km radius, we arrive at  $\Omega \approx 9.5\times 10^3\ \mathrm{s}^{-1}$. We need to compare this to the expected break-up frequency, $\Omega_K$, which is approximated as 
\begin{equation}
{1\over 2\pi} \Omega_K\ [\mathrm{kHz}] \approx 1.716 \left( {M_0 \over 1.4M_\odot} \right)^{1/2} \left( {R_0\over 10\ \mathrm{km}}\right)^{-3/2} - 0.189
\end{equation}
for the models considered in \citet{Doneva:2013zqa} (with $M_0$ and $R_0$ the mass and radius of the corresponding non-rotating model, respectively). Naively using this estimate for our suggested HMNS parameters, we would have $\Omega_K \approx 2.6\times10^4\ \mathrm{s}^{-1}$. Taking these estimates at face value, the HMNS would rotate at just below half of the Kepler rate. 

Armed with this (rough) estimate, let us turn to the f-mode frequency. Intuitively one would expect the f-mode that co-rotates with the orbit to be the one that is excited by the merger dynamics simply because this mode most closely resembles the configuration when the two stars come into contact.  With the conventions from  \citet{Doneva:2013zqa}, we are then considering the stable $l=-m=2$ mode and, 
in a frame rotating with the star, we have
\begin{equation}\label{f-mode_comov_stable}
\sigma_r \approx \sigma_0 \left[ 1 - 0.235 \left( {\Omega \over \Omega_K}\right) -0.358 \left( {\Omega \over \Omega_K}\right)^2 \right]
\end{equation} 
However, this needs to be translated into the inertial frame (where the gravitational-wave signal is measured). Using $\sigma_i = \sigma_r - m\Omega$ we obtain
\begin{equation}\label{f-mode_inertial_stable}
\sigma_i = \sigma_0 \left[ 1 - 0.235 \left( {\Omega \over \Omega_K}\right)-\frac{m\Omega}{\sigma_0} -0.358 \left( {\Omega\over \Omega_K}\right)^2 \right] 
\end{equation}
where $\sigma_0$ is the f-mode frequency of a non-rotating star. This is estimated as
\begin{equation}
{1\over 2\pi} \sigma_0  [\mathrm{kHz}] \approx 1.562 + 1.151  \left( {M_0 \over 1.4M_\odot} \right)^{1/2} \left( {R_0\over 10\ \mathrm{km}}\right)^{-3/2}
\end{equation}
which for our fiducial parameters returns $\sigma_0 \approx \Omega_K$.
Taking $\Omega/\Omega_K\approx 0.5$ we then have 
\begin{equation}
\sigma_i \approx 1.8 \sigma_0
\end{equation}

We thus arrive at a back-of-the-envelope  idea of how much  the f-mode changes due to rotation, compared to a  rotating model. Basically, we estimate that rotation would take us another factor of almost 2 towards explaining the results in figure~\ref{f2vk}.

\section{Discussion and Concluding Remarks}

Using simple approximations, we have tried to explain universal behaviour seen in simulations of neutron star mergers. The basic premise was that we wanted to demonstrate the intuitive association of the dominant oscillation seen in merger simulations with the fundamental oscillation mode of the HMNS.  In order to argue the case, we compared the inferred f-modes for hot, differentially rotating post-merger HMNSs to robust scalings relevant for cold NSs. We also made use of phenomenological relations for the oscillations of rapidly (and uniformly) rotating neutron stars. Very roughly, our estimates suggest that thermal effects  should increase the f-mode frequency by a factor of (up to) 2, compared to the cold neutron star f-mode. Similarly, rotation would introduce another factor of 2. Combining the effects, the post-merger frequencies should lie  a factor 3-4 above the frequency of the individual pre-merger stars. While admittedly simplistic, this estimate allows us to connect the observed frequency HMNS to the (much more easily calculated) f-modes of cold single neutron star. This would ``explain'' why the features seen in simulations are found to be robust.

The arguments we have provided are obviously at the level of back-of-the-envelope estimates. Nevertheless, the exercise opens the door for future possibilities. First, one should be able to extend the rotational estimates to the realistic case of differential rotation (see, for example, \cite{uryu} for the differential rotation profiles expected for post-merger remnants). This would be an important step as it should quantify the rotational effects. Moreover, given the available computational technology \citep{Doneva:2013zqa}, such results should be within reach.  The issue of thermal effects (and potential phase transitions, \citet{han}) is more complex, as aspects related to heat (entropy) tend to be treated in a somewhat ad hoc manner in many numerical simulations. Ultimately, one would like demonstrate that the observed relation between tidal deformability and post-merger dynamics is real and not an artefact due to (for example) a ``simplified'' treatment of the physics. There is scope for improvements in this direction, although it will require some effort as it involves paying detailed attention to the thermodynamics in nonlinear simulations. 

\section*{Acknowledgements}
We would like to, first of all, thank Sebastiano Bernuzzi for providing the numerical simulation data used in figure~\ref{f2vk} and acknowledge the use of data from the ComPOSE website for the thermal EoSs.
NA gratefully acknowledges support from  STFC via grant ST/R00045X/1. Research of K.C. was supported in part by the International Centre for Theoretical Sciences (ICTS) during a visit for participating in the program Summer School on Gravitational-Wave Astronomy (Code: ICTS/Prog-gws/2017/07).

\bibliographystyle{mn2e}

\end{document}